\documentclass{elsart}
\usepackage[dvips]{graphicx}
\usepackage{amssymb}

\begin{document}

\begin{frontmatter}

    \title{A Comparison of High-Frequency Cross-Correlation Measures}
    \author[London]{Ovidiu V. Precup},
    \author[London]{Giulia Iori}
    \address[London]{Mathematics Department, King's College, University of London, Strand, London WC2R 2LS, UK}
    \vspace{-3mm}
\begin{abstract}
 On a high-frequency scale the time series are not homogeneous, therefore standard correlation measures can not
be directly applied to the raw data. There are two ways to deal with this problem. The time series can be homogenised
through an interpolation method~\cite{Dacorogna} (linear or previous tick) and then the Pearson correlation statistic
computed. Recently, methods that can handle raw non-synchronous time series have been developed~\cite{Reno1,deJong}.
This paper compares two traditional methods that use interpolation with an alternative method applied directly to the
actual time series.

\end{abstract}
\end{frontmatter}

\section{Introduction}
\vspace{-6mm}
 In this paper we present and compare three different methods of computing the cross correlation matrix
from high frequency equity trades data. The component stocks of the S\&P100 are employed in analysing the NYSE Trades
and Quotes (TAQ) database.  In the context of this paper, high-frequency data is defined as the raw time series of
trades. The time interval between transactions ranges from zero seconds (several distinct trades recorded at the same
time) to forty minutes.\newline
\indent An extension of the standard Pearson correlation measure is proposed in
\cite{Dacorogna} by incorporating a ''covolatility weighting`` for the time series. The weight has the role of
emphasizing periods where trading has a noticeable effect on asset prices. \newline \indent Let $ X, Y $ be two asset
price time series which have been homogenised and synchronised to a time step $\Delta t$, covolatility weights are
given by $\omega_i$ and time length of the trading period is $T$. We define $\Delta x$, $\Delta y$ as the corresponding
log returns series on a time scale $\Delta t$ and $\Delta X$, $\Delta Y$ as the log returns on a larger time scale $m
\Delta t$. The covolatility adjusted correlation measure is defined as: \vspace{-1mm}
\begin{equation}\label{CovCorr}
\rho(\Delta X_i, \Delta Y_i)=\frac{\sum_{i=1}^{T/m\Delta t}(\Delta X_i-\overline{\Delta X})(\Delta Y_i-\overline{\Delta Y})
\omega_i}{\sqrt{\sum_{i=1}^{T/m\Delta t}(\Delta X_i-\overline{\Delta X})^{2}\omega_i\sum_{i=1}^{T/m\Delta t}
(\Delta Y_i-\overline{\Delta Y})^{2}\omega_i}}
\end{equation}
\vspace{-6mm}
\begin{equation}\label{CovWeight}
\mathrm{where}\quad \omega_i=\sum_{j=1}^{m}(|\Delta x_{i\cdot m-j}-\overline{\Delta x}_{i\cdot m}|\cdot|\Delta y_{i\cdot m-j}-\overline{\Delta y}_{i\cdot m}|)^{\alpha},
\end{equation}
\vspace{-6mm}
\begin{equation}
\overline{\Delta x}_{i\cdot m}=\sum_{j=1}^{m}\frac{\Delta x_{i\cdot m-j}}{m}, \quad
\Delta X_i=\sum_{j=1}^{m}\Delta x_{i\cdot m-j}, \quad
\overline{\Delta X}=\frac{\sum_{i=1}^{T/m\Delta t}\Delta X_i\cdot \omega_i}{\sum_{i=1}^{T/m\Delta t}\omega_i} .
\end{equation}

\vspace{-4mm}
Setting $\omega_i=1$ reduces (\ref{CovCorr}) to the standard Pearson coefficient. In this paper as in~\cite{Dacorogna}
$\alpha$ = 0.5 but this can be varied so that more weight is given to periods where the returns volatility is above
average. In~\cite{Dacorogna} $m$ = 6, in our analysis it varies from 3 to 480 (the number of time units of $\Delta t$
in the trading day). This was determined by the choice of $\Delta t=60$ seconds which was taken as a tradeoff value
for the average trading interval pattern. The intention is to avoid extensive imputation towards the end of the trading
day when there are few transactions occurring.\newline
\indent
Methods that can be directly applied to the actual time series to obtain correlation statistics have been presented in
\cite{Reno1,Reno2,deJong}. The method by de Jong~\cite{deJong} is based on a regression type estimator but
it relies on a rather strong assumption of independence between prices and transaction times. Barucci and Ren\`o
~\cite{Reno1,Reno2} have adapted a Fourier method developed by Malliavin~\cite{Malliavin} to the computation of FX rates correlations.
The Fourier method is model independent, it produces very accurate, smooth estimates and handles the time series in their
original form without imputation or discarding of data. A rigorous proof of the method is given in the original paper
by Malliavin~\cite{Malliavin} so only the main results are given below.\newline
\indent
Let $S_i(t)$ be the price of asset $i$ at time $t$ and $p_i(t)=\ln S_i(t)$. The physical time interval of the asset price
series is rescaled to $[0,2\pi]$. The variance/covariance matrix $\Sigma_{ij}$ of log returns is derived from its Fourier
coefficient $a_0(\Sigma_{ij})$ which is obtained from the Fourier coefficients of $dp_i$:
\vspace{-2mm}
\begin{equation}\label{FCoeff1}
a_k(dp_i)=\frac{1}{\pi}\int_{0}^{2\pi} \cos(kt)dp_i(t),\quad
b_k(dp_i)=\frac{1}{\pi}\int_{0}^{2\pi}\sin(kt)dp_i(t),\quad \scriptstyle k\geq 1 .
\end{equation}
\vspace{-3mm}
In practice, the coefficients are computed through integration by parts:
\begin{displaymath}
a_k(dp_i)=\frac{p(2\pi)-p(0)}{\pi}+\frac{k}{\pi}\int_{0}^{2\pi}\sin(kt)p_i(t)dt,\quad
b_k(dp_i)=-\frac{k}{\pi}\int_{0}^{2\pi} \cos(kt)p_i(t)dt .
\end{displaymath}
\vspace{-3mm}
The Fourier coefficient of the pointwise variance/covariance matrix $\Sigma_{ij}$ is :
\begin{equation}\label{FCov1}
a_0(\Sigma_{ij})=\lim_{\tau \rightarrow 0}\frac{\pi\tau}{T}\sum_{k=1}^{T/2\tau}[a_k(dp_i)a_k(dp_j)+
b_k(dp_i)b_k(dp_j)] .
\end{equation}
\vspace{-4mm} The smallest wavelength ($T/2\tau$) that can be analysed before encountering aliasing effects is
determined by the lower bound of $\tau$ (time gap between two consecutive trades) which is 1 second for all S\&P100
price series. The integrated value of $\Sigma_{ij}$ over the time window is defined as
\mbox{$\hat{\sigma}^{2}_{ij}=2\pi a_0(\Sigma_{ij})$} which leads to the Fourier correlation matrix
$\rho_{ij}={\hat{\sigma}^{2}_{ij}}/({\hat{\sigma}_{ii}\cdotp\hat{\sigma}_{jj}})$.
\vspace{-5mm}
\section{Results}
\vspace{-3mm} We tested the Fourier method on simulated bivariate GARCH(1,1) processes in a similar setting to that
in~\cite{Reno1}. The time interval $\delta$ between trades in S\&P100 equities approximately follows an exponential
distribution with rate parameter $\beta$ in the range 1 (very liquid stock) to 22 (least liquid stock) seconds. We
sampled the generated GARCH process using the exponential distribution and varied $\beta$ so as to resemble actual
trading patterns.\newline \indent The method works very well on synchronous series with random gaps irrespective of the
rate $\beta$. The tests on asynchronous series with $\beta\leq 6$ were also successful for the entire correlation
spectrum. When $\beta\geq 10$ for at least one of the two series, the correlation decays noticeably at time scales
smaller than 5 minutes but converges quickly to the induced value when the time scale is greater than 10 minutes. The
correlation decay in the high-frequency regime seems to be directly related to the rate parameter $\beta$.

\begin{figure}[h]
    \includegraphics[scale=0.4]{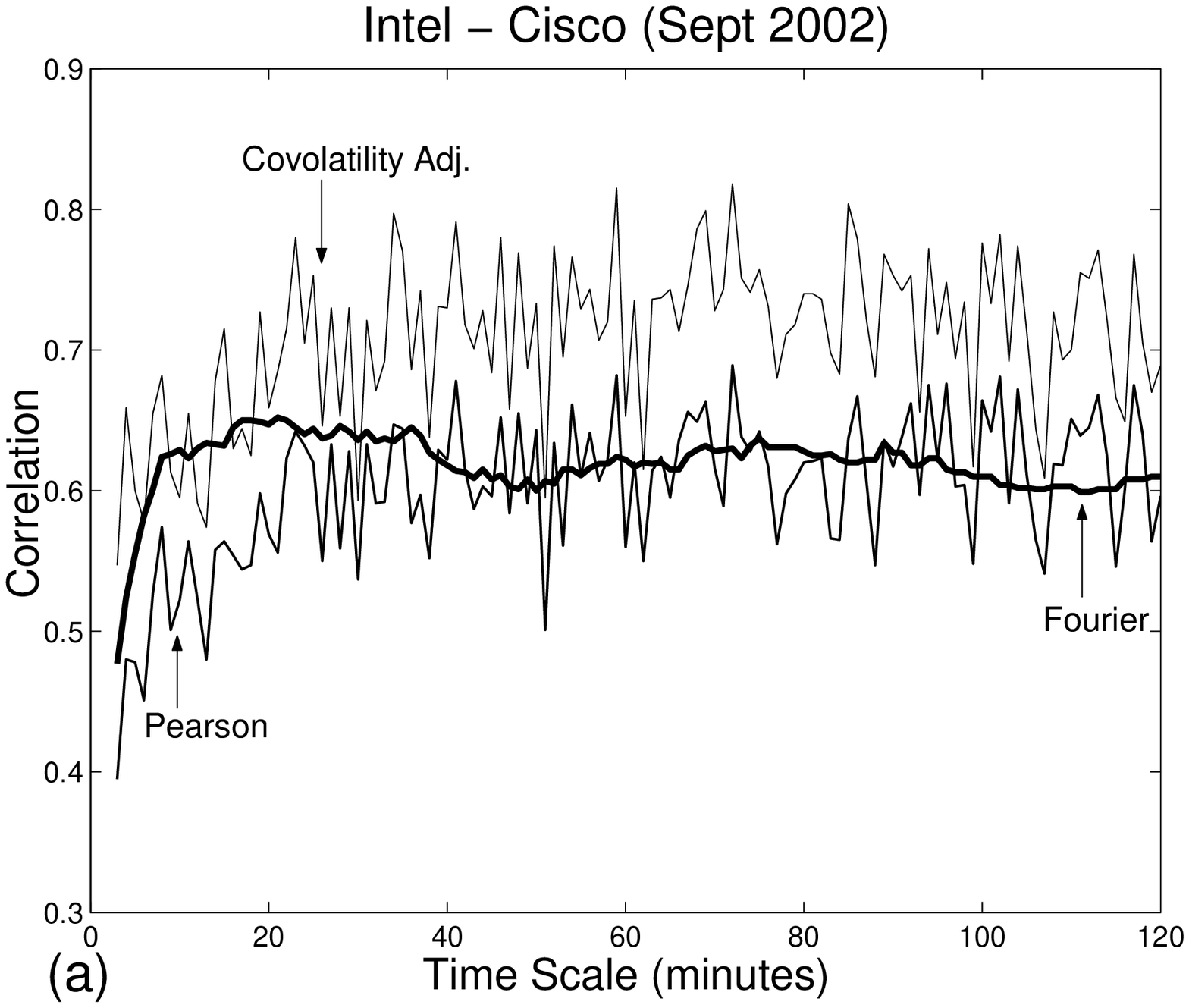}\quad
    \includegraphics[scale=0.4]{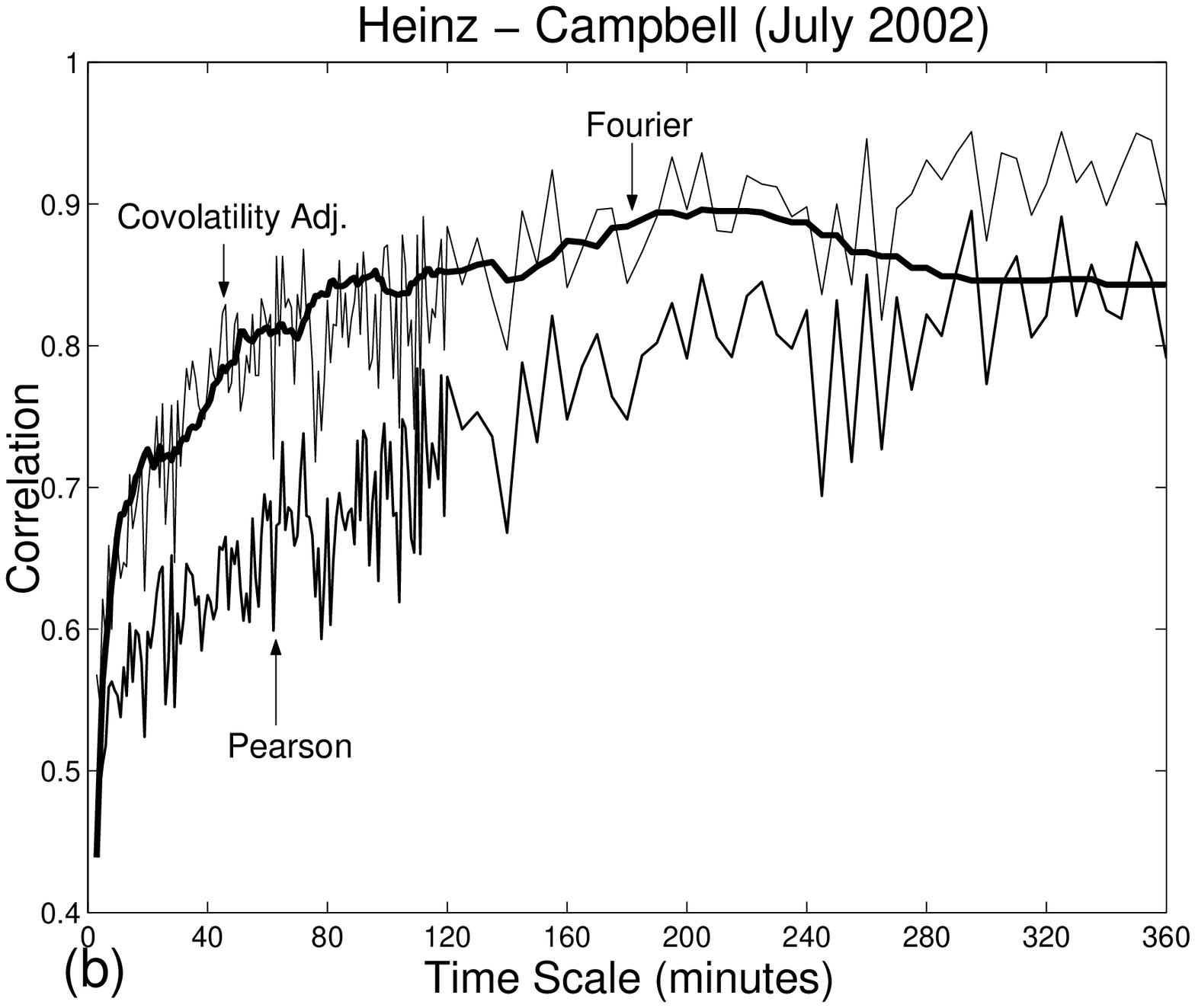}
    \caption{\scriptsize Time scale values correspond to $m$ for Pearson and Covolatility Adjusted methods and
    $\tau$ for Fourier method. The rate parameters are $\beta_{intel}$=1, $\beta_{cisco}$=1.2 in Fig.1.(a),
    $\beta_{heinz}$=13, $\beta_{campbell}$=21 in Fig.1.(b).}
\end{figure}

In Fig.1 the correlation spectra for two pairs of stocks is shown computed with each of the three correlation measures.
In both cases the Fourier correlation method provides a much smoother spectrum than the other two methods (Pearson and
Covolatility Adjusted) which use interpolation. The ''Epps effect``\footnote{The correlation between stocks falls when
decreasing the time scale~\cite{Epps}.} can also be observed in the two plots and displays one of the properties
described in~\cite{Lundin}, the more an asset is traded, the less marked the Epps effect is. The correlation between
Intel and Cisco (highly liquid, $\beta_{intel}$=1, $\beta_{cisco}$=1.2) reaches a stable level after approximately 15
minutes whilst for Heinz and Campbell (lower liquidity, $\beta_{heinz}$=13, $\beta_{campbell}$=21) it takes about 2
hours to stabilise.
\newline
\indent The Fourier method averages the Covolatility adjusted correlation measure at very high frequency (time scale
under 20 minutes) and trails a moving average of the Pearson coefficient at lower frequencies. This indicates that the
method is also robust in the low frequency domain where the Pearson method can be taken as the benchmark. The
correlation decay observed in the GARCH tests at time scales of less than 5 minutes due to non-synchronicity in trades
can not account for the correlation structure that develops over 2 hours in Fig.1.(b). Thus, the Epps effect present in
the correlation structure of illiquid stocks can not be explained by non-synchronicity in transactions but is an actual
market microstructure phenomenon related to the information aggregation and price formation processes. From the
analysis carried out it can be inferred that the Fourier method of computing the correlation matrix from high-frequency
data is better than the alternatives in terms of generating smooth, robust estimates. It is conceptually superior to
methods that use interpolation and is also model independent. Further studies are under way~\cite{Precup} to explore
other contributing factors to the Epps effect, the impact of trading synchronicity on the correlation measure and the
time-scale dynamics of correlation matrices.

\vspace{-6mm}
\section{Acknowledgements}
\vspace{-6.5mm} We are very grateful to Michel Dacorogna and Roberto Ren\`o for enlightening discussions. 


\vspace{-5mm}
\begin{thebibliography}{99}
\vspace{-5mm}
    \bibitem{Dacorogna} M. Dacorogna, R. Gencay, U.A. M\"uler, R.B. Olsen, O.V. Pictet, An Introduction to High-Frequency
            Finance (2001), Academic Press.
\vspace{-3mm}
    \bibitem{Reno1} R. Ren\`o, A closer look at the Epps effect, International Journal of Theoretical and
            Applied Finance, (2003), 6 (1), 87-102.
\vspace{-3mm}
    \bibitem{Reno2} E. Barucci, R. Ren\`o, On measuring volatility and the GARCH forecasting performance, Journal of
            International Financial Markets, Institutions and Money, 12 (2002) 182-200.
\vspace{-3mm}
    \bibitem{deJong} F. de Jong, T. Nijman, High frequency analysis of lead-lag relationships between financial markets,
            Journal of Empirical Finance 4 (1997) 259-277.
\vspace{-3mm}
    \bibitem{Malliavin} P. Malliavin, M. Mancino, Fourier series method for measurement of multivariate volatilities,
            Finance \& Stochastics 6(1), (2002), 49-61.
\vspace{-3mm}
    \bibitem{Epps} T. Epps, Comovements in stock prices in the very short run, Journal of the American Statistical
            Association 74, (1979), 291-298.
\vspace{-3mm}
    \bibitem{Lundin} M. Lundin, M. Dacorogna, Correlation of high-frequency financial time series, in P. Lequeux (Ed.),
            Financial Markets Tick by Tick (1999), Wiley \& Sons.
\vspace{-3mm}
    \bibitem{Precup} O.V. Precup, G. Iori, High-Frequency Cross-Correlation Dynamics in US Equity Markets, (in preparation).
\end{thebibliography}
\end{document}